\documentclass[graybox,natbib,nosecnum]{svmult}
\bibpunct{(}{)}{;}{a}{}{,} % suppress commas between author-names and year

\pdfoutput=1   %forces use of pdflatex. Disable if you prefer to use .eps or .ps figures.
\usepackage[varg]{txfonts}       % \lesssim
\usepackage{mathptmx}       % selects Times Roman as basic font
\usepackage{helvet}         % selects Helvetica as sans-serif font
\usepackage{courier}        % selects Courier as typewriter font
\usepackage{makeidx}         % allows index generation
\usepackage{graphicx}        % standard LaTeX graphics tool
\usepackage{multicol}        % used for the two-column index
\usepackage[bottom]{footmisc}% places footnotes at page bottom
\usepackage[normalem]{ulem}	% for strike-through of text with \sout{}
\usepackage{hyperref}  %for hyperlinks
\hypersetup{colorlinks=true, citecolor=blue, linkcolor=blue}

\usepackage{soul}   % for high-lighting of text
\newcommand*\hbindex{}  %highlights index entries

\newcommand*\mD{D}
\newcommand*\mH{H}
\newcommand*\mK{K}
\newcommand*\mR{\tilde R}
\newcommand*\mS{\tilde S}

\def\ImUnit{\mathbf{i}}

\def\expo#1{\mathbf{e}^{#1}}

\def\scaled#1{\hat{#1}}

\def\figpath{}

\makeindex             % used for the subject index
\begin{document}

%\title*{Planets in mean-motion resonances: Gravitational interactions %and orbital resonances
%between planets} %extrasolar planets}
\title*{Planets in Mean-Motion Resonances and the System Around HD45364} 
\titlerunning{Planets in mean-motion resonances}
\author{Alexandre C. M. Correia, Jean-Baptiste Delisle, and Jacques Laskar}
\institute{Alexandre C. M. Correia \at CIDMA, Departamento de F\'isica, Universidade de Aveiro, Campus de Santiago, 3810-193 Aveiro, Portugal, \email{correia@ua.pt}
\and Jean-Baptiste Delisle \at Observatoire de l'Universit\'e de Gen\`eve, 51 chemin des Maillettes, 1290 Sauverny, Switzerland, \email{jean-baptiste.delisle@unige.ch}
\and Jacques Laskar \at ASD, IMCCE-CNRS UMR8028, Observatoire de Paris, 77 Av. Denfert-Rochereau, 75014 Paris, France, \email{Jacques.Laskar@obspm.fr}}
\maketitle

\abstract{In some planetary systems the orbital periods of two of its members present a commensurability, usually known by mean-motion resonance.
These resonances greatly enhance the mutual gravitational influence of the planets.
As a consequence, these systems present uncommon behaviours and their motions need to be studied with specific methods.
Some features are unique and allow us a better understanding and characterisation of these systems.
Moreover, mean-motion resonances are a result of an early migration of the orbits in an accretion disk, so it is possible to derive constraints on their formation.
Here we review the dynamics of a pair of resonant planets and explain how their orbits evolve in time.
We apply our results to the HD\,45365 planetary system.
}

\section{Introduction}
In addition to the solar system, about 600 multi-planetary systems are known, that is, systems that contain at least two planets.
Among  these systems, at least eleven pairs of planets have been identified to have resonant orbits, or, more precisely, to be trapped in a \hbindex{mean-motion resonance} (see Table~\ref{tab1}).
Many other pairs are listed as possible resonances, but our present knowledge of their orbital elements is not sufficient to confirm the resonant behaviour.

A mean-motion resonance occurs when two planets exert a regular, periodic gravitational influence on each other.
The physics principle behind is similar in concept to a driven harmonic oscillator.
As for the oscillator, a planet has a natural orbital frequency. 
Another planet doing the ``pushing" will act in periodic repetition and has a cumulative effect on the motion of the considered planet.
In mean-motion resonances, the orbital periods are  related by a ratio of two small integers:
\begin{equation}
\frac{P_2}{P_1} \approx \frac{p+q}{p} \ , \label{161122a}
\end{equation}
where $P_1$ and $P_2$ are the orbital periods of the inner and outer planet, respectively, and $p$ and $q$ are integers.
The $q$ value also designates the order of the resonance, and the resonance is called a $(p+q)$:$p$ mean-motion resonance.
Mean-motion resonances greatly enhance the mutual \hbindex{gravitational interactions} between the planets. %, i.e., their ability to alter or constrain each other's orbits.
In many cases, the result is an unstable interaction, in which the planets exchange momentum and shift orbits until the resonance no longer exists.
For instance, the Kirkwood gaps in the spatial distribution of the asteroids correspond to orbital periods that are integer fractions of Jupiter's period.
However, under some circumstances, mean-motion resonances can be self-correcting, so that the system remains stable with the planets in resonance.
This is the case for all the planetary systems observed in mean-motion resonances (Table~\ref{tab1}).

\begin{table}
    \caption{Extra-solar planets confirmed to be trapped in a \hbindex{mean-motion resonance}.}
    \label{tab1}
    \begin{tabular}{p{2cm}p{2cm}p{2cm}p{4cm}}
      \hline
      Star & Planets & Resonance & Reference\\
      \hline
      GJ 876    &  c,b  & 2$:$1 & \citet{Lee_Peale_2002}\\
      GJ 876    &  b,e  & 2$:$1 & \citet{Rivera_etal_2010}\\
      HD 73526  &  b,c  & 2$:$1 & \citet{Tinney_etal_2006}\\
      HD 82943  &  c,b  & 2$:$1 & \citet{Lee_etal_2006}\\
      HD 128311 &  b,c  & 2$:$1 & \citet{Vogt_etal_2005}\\
      HD 45364  &  b,c  & 3$:$2 & \citet{Correia_etal_2009}\\
      HD 204313 &  b,d  & 3$:$2 & \citet{Robertson_etal_2012}\\
      HD 5319   &  b,c  & 4$:$3 & \citet{Giguere_etal_2015}\\
      HD 60532  &  b,c  & 3$:$1 & \citet{Laskar_Correia_2009}\\
      HD 33844  &  b,c  & 5$:$3 & \citet{Wittenmyer_etal_2016}\\
      HD 202206 &  b,c  & 5$:$1 & \citet{Correia_etal_2005}\\
      \hline
    \end{tabular}
\end{table}

\begin{figure}
  \centering
    \includegraphics*[width=\linewidth]{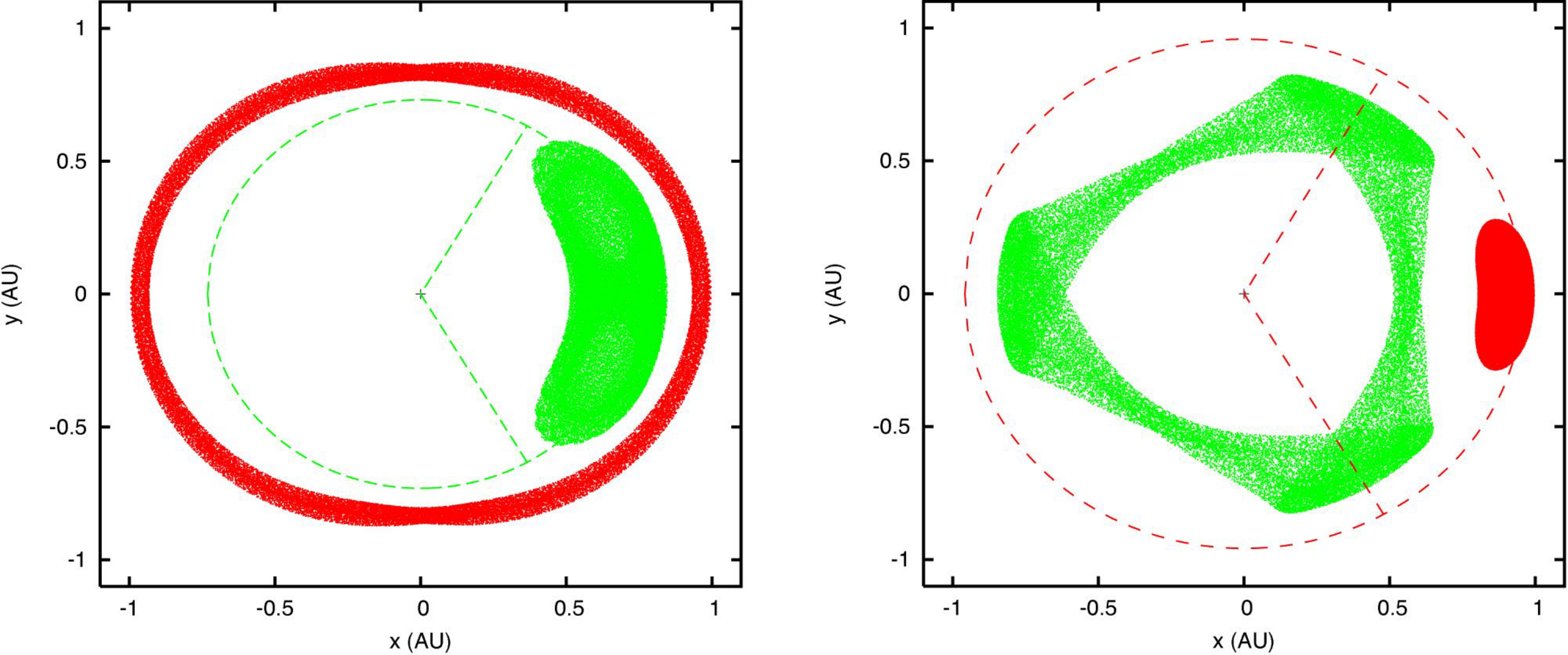}
  \caption{Evolution of the HD\,45364 planetary system over $10^5$~yr in the
  co-rotating frame of the inner planet, $b$ ({\it left}) and outer planet, $c$ ({\it right}). $x$ and
  $y$ are spatial coordinates in a frame centered on the star and rotating
  with the frequency $n_b$ ({\it left}) or $n_c$ ({\it right}).
  Due to the 3:2 mean-motion resonance the planets never get too close, the minimal distance of approach
  being 0.371~AU, while the maximum distance can reach 1.811~AU. We also observe
  that the trajectories are repeated every 3 orbits of the inner planet and
  every 2 orbits of the outer planet.
  In this particular frame, we are also able to see the libration of each planet around its equilibrium position.}
\label{F5}
\end{figure}

In the solar system, we observe a 3:2 mean-motion resonance between Neptune and Pluto, which means that Pluto completes two orbits in the time it takes Neptune to complete three.
Another example of a 3:2 resonance is present in the \hbindex{HD\,45364} planetary system, but involving two giant planets \citep{Correia_etal_2009}.
In Fig.\,\ref{F5}, we show the evolution of the HD\,45364 system over $10^5$~yr in the rotating frame of the inner and of the outer planet.
Due to the 3:2 mean-motion resonance trapping, the relative positions of the two planets are repeated, preventing close encounters and the consequent destruction of the system.
The paths of the planets in the rotating frame illustrate the relationship between the resonance and the frequency of conjunctions with the internal or external planet: the orbital configuration of the system is repeated every 3 orbits of the inner planet and every 2 orbits of the outer planet.
In this particular frame, we are also able to see the libration of each planet around its equilibrium position.

The most commonly observed mean-motion resonance is the 2:1, followed by the 3:2, but many other configurations are also possible (Table~\ref{tab1}).
Sometimes more than two planets can be involved in a mean-motion resonance.
In the solar system this is the case of the Jupiter's moons Io, Europa and Ganymede, which are involved in a 4:2:1 resonance, also known as {\it Laplace resonance}.
For exoplanets, a similar example is known for the \hbindex{GJ\,876} planets $b$, $c$ and $e$ \citep{Rivera_etal_2010}.

%The physics principle behind mean-motion resonances is similar in concept to a driven harmonic oscillator, where one of the orbits and the oscillator both have a natural frequency, and the other planet doing the ``pushing" will act in periodic repetition to have a cumulative effect on the motion.
%For simplicity, consider to planets with masses $m_1 \ll m_2$, whose orbital periods are commensurable (Eq.\,\ref{161122a}).
%During the period from opposition to conjunction, $m_2$ removes angular momentum from $m_1$ via the tangential component of the perturbing mutual force, and from conjunction to opposition it adds angular momentum.
%If the conjunction occurs exactly at pericenter or apocenter, these effets integrate to zero, and there is no net transfer.
%However, repetitive conjunctions at any other point destroy this symmetry...
%For more details see \citet{Peale_1976}.

\section{Conservative dynamics}
\label{sec:conservative}

We refer to the star as body 0, to the inner planet as body 1, and to the outer planet as body 2.
Denoting the masses of the three bodies $m_i$, we introduce for both planets
$\mu_i = \mathcal{G} (m_0 + m_i)$
and $\beta_i = m_0 m_i/(m_0 + m_i)$,
where $\mathcal{G}$ is the gravitational constant.
We let $\mathbf{r}_i$ be the position vectors of the planets with respect to the star and
$\tilde{\mathbf{r}}_i$ the canonically conjugated momenta
\citep[in astrocentric coordinates, see][]{Laskar_Robutel_1995}.
As usual in the literature, semi-major axes are noted $a_i$, eccentricities $e_i$,
mean longitudes $\lambda_i$, and longitudes of periastron $\varpi_i$.
For simplicity, we only consider the planar case.
The Hamiltonian of the three-body problem reads
\begin{equation}
  \label{eq:H3b}
  \scaled{\mH} = \scaled{\mK} + \scaled{\mH}_1\ ,
\end{equation}
where $\scaled{\mK}$ is the Keplerian part (star-planets interactions)
and $\scaled{\mH}_1$ is the perturbative part (planet-planet interactions).
The Keplerian part is given by
\begin{equation}
  \label{eq:Kep}
  \scaled{\mK} = - \sum_{i=1}^2 \frac{\mu_i^2 \beta_i^3}{2 \scaled{\Lambda}_i^2}\ ,
\end{equation}
where $\scaled{\Lambda}_i = \beta_i \sqrt{\mu_i a_i}$ is the circular angular momentum of planet $i$.

The perturbative part can be decomposed in direct and indirect interactions,
\begin{equation}
  \scaled{\mH}_1 =-\mathcal{G} \frac{m_1m_2}{\Delta_{12}}
+ \frac{\mathbf{\tilde{r}}_1 \cdot \mathbf{\tilde{r}}_2}{m_0}    \label{eq:perturb} \ ,
\end{equation}
%\begin{eqnarray}
%  \label{eq:perturb}
%  \scaled{\mH}_1 &=& \scaled{\mU}_1 + \scaled{\mT}_1\\
%  \scaled{\mU}_1 &=& -\mathcal{G} \frac{m_1m_2}{\Delta_{12}}\\
%  \scaled{\mT}_1 &=& \frac{\mathbf{\tilde{r}}_1 . \mathbf{\tilde{r}}_2}{m_0}\ ,
%\end{eqnarray}
with $\Delta_{12} = ||\mathbf{r}_1-\mathbf{r}_2||$.
This perturbation can be expressed as a function of elliptical orbital elements by expanding it
in Fourier series of the mean longitudes $\lambda_i$, and longitudes of the pericenter $\varpi_i$ \citep[e.g.,][]{Laskar_Robutel_1995}.
For a given mean-motion resonance $(p+q)$:$p$, the corresponding
combination of the mean longitudes undergoes slow variations (Eq.\,(\ref{161122a}))
\begin{equation}
  (p+q) n_2 - p n_1 \approx 0\ ,
\end{equation}
where $n_i = \dot \lambda_i \approx 2 \pi /P_i$,
while other (non-resonant) combinations of these angles circulate rapidly.
The long-term evolution of the orbits is thus accurately described by the averaged Hamiltonian
over the non-resonant combinations of the mean longitudes.
By performing this averaging and the classical angular momentum reduction,
one obtains a two degrees of freedom problem \citep[e.g.,][]{Delisle_etal_2012} and
two constants of motion.
The first constant of motion is the total \hbindex{angular momentum},
\begin{equation}
  \label{eq:G}
  \scaled{G} = \scaled{G}_1 + \scaled{G}_2 = \scaled{\Lambda}_1\sqrt{1-e_1^2}
  + \scaled{\Lambda}_2\sqrt{1-e_2^2}\ .
\end{equation}
The second, which comes from the averaging, is a combination of the circular angular momenta
\citep[or semi-major axes, e.g.,][]{Michtchenko_Ferraz-Mello_2001},
\begin{equation}
  \label{eq:Gamma}
  \Gamma = \frac{p+q}{p} \scaled{\Lambda}_1 + \scaled{\Lambda}_2\ .
\end{equation}
As shown in \citet{Delisle_etal_2012}, the constant $\Gamma$ can be used as a scaling
factor and does not influence the dynamics of the system
except by changing the scales of the problem (in space, energy, and time).
The elimination of $\Gamma$ is achieved by performing the following change of coordinates:
$\Lambda_i = \scaled{\Lambda}_i / \Gamma$,
$G_i = \scaled{G}_i / \Gamma$,
$\mH = \Gamma^2 \scaled{\mH}$, and
$t = \scaled{t}/\Gamma^3$,
while angle coordinates are unchanged.
Using these new coordinates, the dynamics of the system depends on only one parameter,
$G = \scaled{G} / \Gamma = G_1 + G_2$, the renormalized angular momentum.
The remaining two degrees of freedom can be represented by both \hbindex{resonant angles},
\begin{equation}
  \label{eq:sigma}
  \sigma_i = -\frac{p}{q}\lambda_1 + \frac{p+q}{q} \lambda_2 - \varpi_i\ , \quad i=1,2,
\end{equation}
and both angular momentum deficits \citep{Laskar_2000},
which are canonically conjugated to the resonant angles,
\begin{equation}
  \label{eq:AMDi}
  I_i = \Lambda_i - G_i = \frac12 \Lambda_i \, \xi_i^2 \propto e_i^2 \ , \quad i=1,2,
\end{equation}
with
\begin{equation}
  \label{eq:xi}
  \xi_i = \sqrt{2\left(1-\sqrt{1-e_i^2}\right)} \approx e_i.
\end{equation}
We can also introduce rectangular coordinates,
\begin{equation}
  \label{eq:bxi}
  x_i = \sqrt{I}_i \expo{\ImUnit \sigma_i}\ .
\end{equation}
It should be noted that for small eccentricities, $|x_i| \propto e_i$.
The averaged Hamiltonian takes the form
\begin{equation}
  \label{eq:Haver}
  \mH = \mK(I_i) + \mS(I_i, \Delta\varpi)
  + \mR(I_i,\sigma_i)\ ,
\end{equation}
where $\mS$ is the secular part of the Hamiltonian depending on the difference
of longitudes of periastron ($\Delta\varpi=\sigma_2-\sigma_1$),
but not on mean longitudes of the planets,
and $\mR$ is the resonant part.
These two parts can be expanded as power series of eccentricities, or, more precisely, of $x_i$
\citep{Laskar_Robutel_1995,Delisle_etal_2012}.

The Keplerian part can be expressed as a function of the momenta $I_i$ by substituting
the expressions of $\Lambda_i$ in Equation~(\ref{eq:Kep}),
\begin{equation}
  \label{eq:La12}
  \Lambda_1 = \Lambda_{1,0} - \frac{p}{q} \left(\mD-\delta\right) \ , \quad
  \Lambda_2 = \Lambda_{2,0} + \frac{p+q}{q} \left(\mD -\delta\right) \ ,
\end{equation}
where $\mD$ is the total \hbindex{angular momentum deficit} \citep{Laskar_2000},
\begin{equation}
  \mD = I_1 + I_2 = \Lambda_1-G_1 + \Lambda_2-G_2 = \sum_{i=1}^2 x_i\bar{x}_i\ ,
\end{equation}
$\delta$ is the total angular momentum deficit at exact commensurability
(which is a constant of motion),
\begin{equation}
  \label{eq:delta}
  \delta = \Lambda_{1,0}-G_1 + \Lambda_{2,0}-G_2 \ ,
\end{equation}
and $\Lambda_{i,0}$ denotes the value of $\Lambda_i$ at exact commensurability.

The secular part contains terms of degree two and more in eccentricities while the resonant part
contains terms of degree $q$ and more.
Thus, the simplest model of the resonance should take into account
at least those terms of order $q$ in eccentricities in the perturbative part,
\begin{equation}
  \label{eq:Hq}
  \mH = \mK(\mD) + \mS_q(I_i, \Delta\varpi)
  + \sum_{k=0}^q R_k (x_1^kx_2^{q-k} + \bar{x}_1^k\bar{x}_2^{q-k})\ ,
\end{equation}
where $\mS_q$ is the secular part truncated at degree $q$, and $R_k$ are
constant coefficients \citep[see][]{Delisle_etal_2012}.
The equations of motion are given by
\begin{equation}
\label{161123b}
\frac{d x_i}{dt} = \ImUnit \, \frac{\partial \mH}{\partial \bar{x}_i} \ .
\end{equation}
This problem is much simpler than the initial four degrees of freedom problem.
However, in general it is still non-integrable since it presents two degrees of freedom.
%\subsection{First-order resonances}
%\label{sec:first-order-reson}
For a first-order mean-motion resonance (such as the 2:1 or 3:2 resonances) the simplest Hamiltonian reads
\begin{equation}
  \label{eq:H1}
  \mH = \mK(\mD) + R_1 (x_1 + \bar{x}_1) + R_0 (x_2 + \bar{x}_2)\ ,
\end{equation}
where there are no secular terms since they only appear at degree two.
It is well known that the Hamiltonian~(\ref{eq:H1}) is integrable
\citep[see][]{Sessin_Ferraz-Mello_1984,Henrard_etal_1986, Wisdom_1986}. %,Batygin_Morbidelli_2013a}.
Introducing $R$ and $\phi$ such that
\begin{eqnarray}
R_1 = R \cos \phi \ , \quad
R_0 = R \sin \phi \ ,
\label{161130a}
\end{eqnarray}
and the new coordinates %$u_1$, $u_2$ such that
\begin{equation}
u_1 = |u_1| \expo{\ImUnit \theta} = x_1 \cos \phi + x_2 \sin \phi \ , \quad
u_2 = -x_1 \sin \phi + x_2 \cos \phi  \ ,
\end{equation}
%with $R_\phi$ the rotation matrix
the Hamiltonian~(\ref{eq:H1}) reads
\begin{equation}
  \label{eq:H1u}
  \mH = \mK(\mD) + R (u_1 + \bar{u}_1) \ ,
  \quad \mathrm{with} \quad
  \mD = \sum_{i=1}^2 u_i\bar{u}_i\ .
\end{equation}
In these coordinates, the Hamiltonian does not depend on the angle associated with $u_2$.
It depends only on the action $u_2\bar{u}_2$, which is thus a new constant of motion of the system.
We are then left with only one degree of freedom ($u_1$, $\bar{u}_1$) and the problem is integrable.
However, if one introduces second-order terms in the Hamiltonian of a first-order resonance,
this simplification no longer occurs.
It does not occur either, in the case of higher order mean-motion resonances, even at the minimal degree of development of the Hamiltonian.
%Moreover, the presence of chaotic motion proves that no reduction to an integrable system is possible \citep[see][]{Wisdom_1986}.
It is nevertheless possible to make some additional hypothesis in order to obtain an integrable system that still captures most of the characteristics of the resonant motion \citep[see][]{Delisle_etal_2014}.

\begin{figure}
    \includegraphics[width=\linewidth]{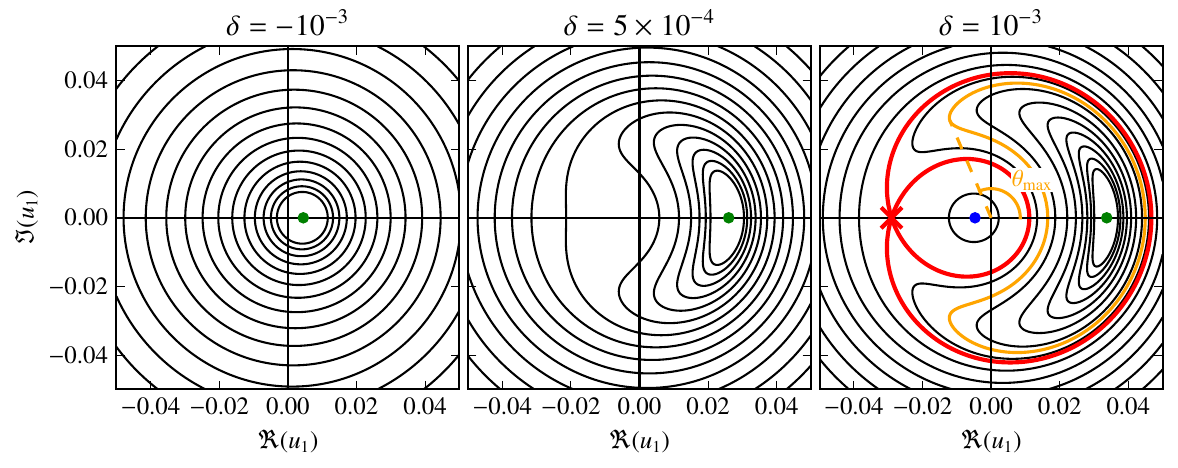}
    \caption{Energy levels sections in the plane defined by $u_2=0$,
    for the 3:2 resonance in the HD\,45364 planetary system for different values of $\delta$.
    Stable fixed points are highlighted with green and blue dots.
    The unstable one is represented by a red cross,
    and the red curves highlight the separatrices of the resonance (\textit{right}).
    In the \textit{right} panel, the blue fixed point is slighlty shifted from zero.
    This is also the case of the green fixed point in the \textit{left} and \textit{center} panels.
    This can cause an artificial libration of the resonant angles $\sigma_i$,
    while the system is not resonant (not in between the separatrices of the resonance).
    }
    \label{Figb}
  \end{figure}

\begin{figure}
    \centering \includegraphics[width=\linewidth]{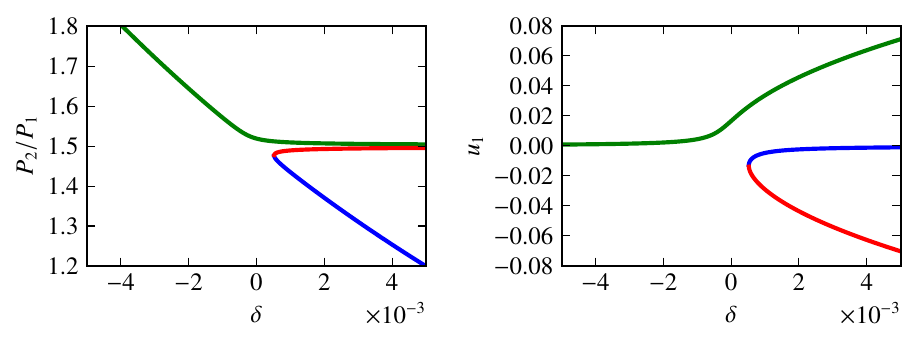}
    \caption{Period ratio (\textit{left}) and value of $u_1$ (\textit{right})
    at the position of the fixed points, as a function of the parameter $\delta$,
    for the 3:2 resonance in the HD\,45364 planetary system.
    All these fixed points lie on the line defined by $u_2=0$ and $\Im(u_1)=0$.
    The colors are the same as in Fig.\,\ref{Figb}.}
    \label{Figg}
  \end{figure}

In Fig.\,\ref{Figb} we show the \hbindex{energy levels} of the Hamiltonian (\ref{eq:H1u})
on the plane defined by $u_2=0$,
in the case of the \hbindex{HD\,45364} planetary system (3:2 mean-motion resonance, $\phi=-31.45^\circ$),
for increasing values of $\delta$ (see Eq.\,(\ref{eq:delta})).
We distinguish three areas of interest: two zones of circulation (internal circulation for low eccentricities and external circulation for high eccentricities), and a libration zone (banana-shaped level curves)
separated from both circulation areas by two separatrices.
\hbindex{Fixed points} of the system, corresponding to $\dot u_i = 0$,
are marked with green and blue dots (for stable ones),
and a red cross (for the unstable one).
The green dot corresponds to the libration center.
In Fig.\,\ref{Figg} we show the positions of these three fixed points
as a function of the parameter $\delta$.
When $\delta$ increases, the libration area moves to higher eccentricities
(i.e. higher values of $|u_1|$, Fig.\,\ref{Figg})
and the internal circulation area takes more space (Fig.\,\ref{Figb}).
On the contrary, for smaller values of $\delta$,
the internal circulation and the libration areas
tend to shrink and eventually completely disappear,
leaving only the external circulation area (Fig.\,\ref{Figb}).

\section{Occurrence of mean-motion resonances}

When multi-planetary systems are found, one may ask whether a pair of its members is in resonance or not.
A straightforward test is to check if the period ratio can be given by the ratio of two integers (Eq.\,(\ref{161122a})).
However, this simple test can be misleading because there is always a rational number close to any real number.
Moreover, the resonant ratio does not depend solely on the orbital periods.
Indeed, form expression (\ref{eq:sigma}), the exact resonance occurs for
\begin{equation}
  \label{161123a}
(p+q) \dot \lambda_2  - p \, \dot \lambda_1 - q \, \dot \varpi_i = 0 \ , \quad i=1,2,
\end{equation}
where $\dot\varpi_i$ corresponds to the precession rate of the apsidal line of planet $i$.
A frequently used test is then to check if the resonant angle $\sigma_i$ (Eq.\,(\ref{eq:sigma})) is in libration.
Nevertheless, for small eccentricities this libration can be artificial \citep{Delisle_etal_2012}: the center of the circulation zone is shifted from zero
(Fig.\,\ref{Figb}),
and the system describes a small circle around the equilibrium point. Thus, 
 the angles $\sigma_i$ appear to librate because their origin is taken at zero
(Fig.\,\ref{Figb}).

The ultimate test to confirm the presence of a mean-motion resonance is to perform a \hbindex{frequency analysis} \citep{Laskar_1988, Laskar_1990} of the orbital solution and check if the mean-motions $n_i$ and the eigenmodes of $\varpi_i$, that we denote $g_i$, are indeed resonant:
\begin{equation}
  \label{161123c}
(p+q) n_2  - p \, n_1 - q \, g_i = 0 \ .
\end{equation}
In addition, unless the \hbindex{libration amplitude} of the resonant angle $\sigma_i$ is fully damped, we should be able to detect a true libration frequency associated with this motion (see Fig.\,\ref{F5}).
In Table\,\ref{T4}, we provide a quasi-periodic decomposition of the resonant angle $ \sigma_1 $ in terms of decreasing amplitude for the 3:2 resonant HD\,45364 planetary system \citep{Correia_etal_2009}.
All the quasi-periodic terms are easily identified as integer combinations of the
fundamental frequencies.
The fact that we are able to express all the main frequencies of $\sigma_1 $ in terms of exact combinations of the fundamental frequencies  is a signature of a very regular resonant motion.

\begin{table}
\caption{Quasi-periodic decomposition of the resonant angle $\sigma_1 = 2 \lambda_1 - 3
  \lambda_2 + \varpi $ for an integration over $10^5$~yr of the orbital solution of the HD\,45364 planetary system \citep{Correia_etal_2009}.}
\label{T4}
\begin{tabular}{p{1cm}p{1cm}p{1cm}p{1cm}p{1cm}p{1.9cm}p{1.9cm}p{1.9cm}}
\hline\noalign{\smallskip}
   \multicolumn{5}{c}{\textbf{Combination}} & \multicolumn{1}{l}{$\nu_i$}
    &\multicolumn{1}{l}{$A_i$}  & \multicolumn{1}{l}{$\phi_i$}  \\
\multicolumn{1}{l}{$n_b$}  & \multicolumn{1}{l}{$n_c$}&
 \multicolumn{1}{l}{$g_1$}  & \multicolumn{1}{l}{$g_2$}
    &\multicolumn{1}{l}{ $l_\sigma$}     & \multicolumn{1}{l}{(deg/yr)}
    &\multicolumn{1}{l}{(deg)}                & \multicolumn{1}{l}{(deg)}  \\
\noalign{\smallskip}\svhline\noalign{\smallskip}
   0 &  0 &  0 &  0 &  1 &    19.8207 &        68.444 &     -144.426 \\
   0 &  0 & -1 & 1 &  0 &     0.8698 &        13.400 &      136.931 \\
   0 &  0 &  1 & -1 &  1 &    18.9509 &         8.606 &      168.643 \\
   0 &  0 & -1 &  1 &  1 &    20.6905 &         8.094 &       82.505 \\
   0 &  0 & -2 &  2 &  0 &     1.7396 &         2.165 &     -176.138 \\
   0 &  0 & -2 &  2 &  1 &    21.5603 &         0.622 &      -50.564 \\
   0 &  0 &  0 &  0 &  3 &    59.4621 &         0.540 &      -73.279 \\
\noalign{\smallskip}\hline\noalign{\smallskip}
\end{tabular}
  We have $\sigma_1 = \sum_{i=1}^N A_i \cos(\nu_i\, t + \phi_i)$, where
  the amplitude and phases $A_i$, $\phi_i$ are given in degree, and the
  frequencies $\nu_i$ in degree/year.
  We only give the first 7 terms, ordered by
  decreasing amplitude. All terms are identified as  integer combinations
  of the fundamental frequencies $n_b$, $n_c$, $g_1$, $g_2$, and the libration frequency $ l_\sigma $.
\end{table}

Since the gravitational interactions for planets involved in mean-motion resonance are very strong, the orbital solutions directly obtained from the observational data are usually subject to large uncertainties.
Indeed, the orbital elements undergo fast variations and therefore models based on Keplerian elliptical orbits are not appropriate.
A correct reduction of the data requires the use of n-body algorithms that take into account the mutual interactions between the planets \citep[e.g.][]{Correia_etal_2005, Correia_etal_2009, Correia_etal_2010, Rivera_etal_2010}.
Even using these algorithms, the best fit solution is sometimes unstable, because small deviations of the true orbital elements render the system chaotic.
A more reliable analysis of resonant systems thus recommends that we perform a global \hbindex{dynamical analysis} in the vicinity of the best fit solution, in order to check for the stable regions.
Only when a planetary system is found inside a stable resonant island we can say with confidence that it is trapped inside a specific mean-motion resonance

\begin{figure}
    \includegraphics*[width=\linewidth]{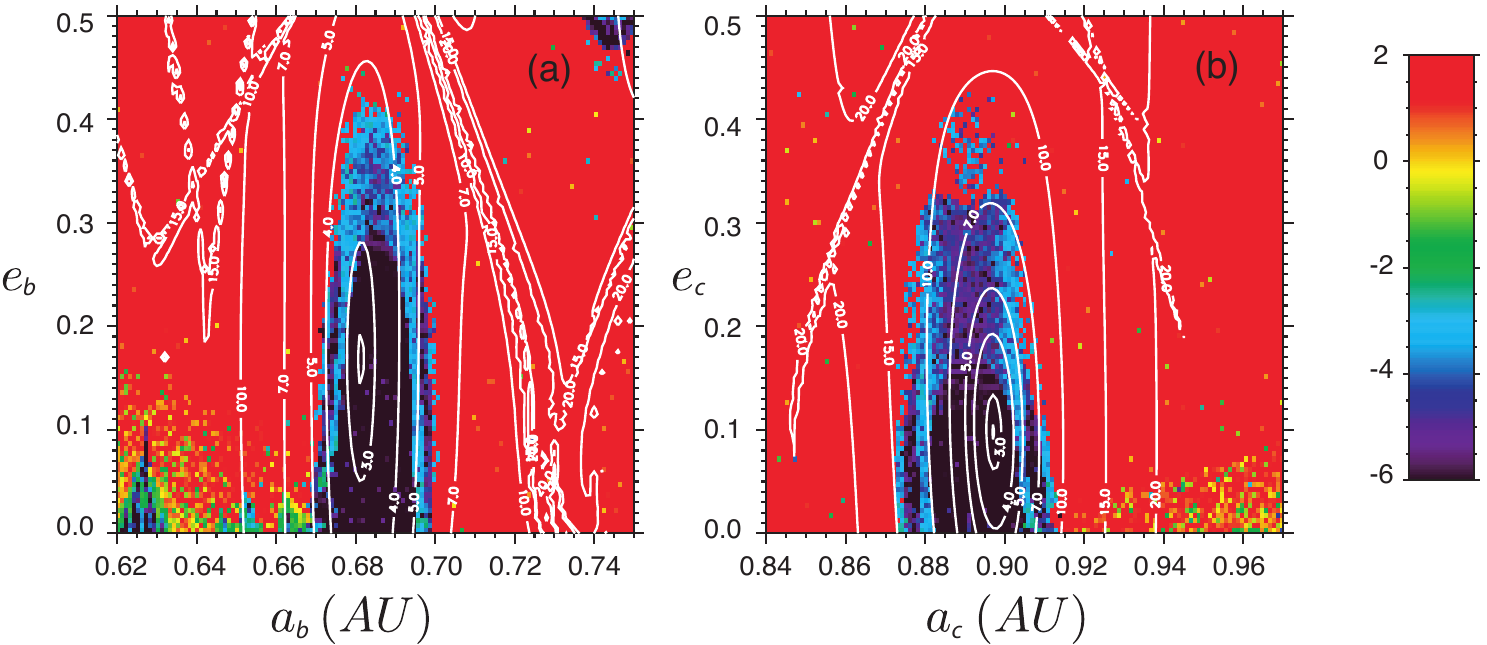}
  \caption{
  Stability analysis of the HD\,45364 planetary system. For a fixed initial condition
  of the outer (a) and inner planet (b), the phase space of the
  system is explored by varying the semi-major axis $a_k$ and eccentricity
  $e_k$ of the other planet, respectively. The step size is $0.001$ AU in semi-major axis
  and $0.005$ in eccentricity. For each initial condition, the
  full system is integrated numerically over $10^4$~yr and a stability criterion
  is derived  with the frequency analysis of the mean longitude
  \citep{Laskar_1990,Laskar_1993PD}.
  As in \citet{Correia_etal_2005}, the chaotic diffusion
  is measured by the variation in the frequencies. The ``red'' zone corresponds to highly unstable
  orbits, while  the ``dark blue'' region can be assumed to be stable on a
  billion-years timescale.
  The contour curves indicate the value of $\chi^2$ %$\sqrt{\chi^2}$
  obtained for each choice of parameters.
  In this case there is perfect
  correspondence between the  zone of minimal $\chi^2$ %$\sqrt{\chi^2}$,
  and the 3:2 stable resonant zone, in ``dark blue''.
  \label{F4}}
\end{figure}

In Fig.\,\ref{F4} we show a \hbindex{stability map} for the HD\,45364 planetary system that confirms the presence of the the 3:2 resonance.
For each planet, the system is integrated on a regular 2D mesh of initial conditions, with varying semi-major axis and eccentricity, while the other parameters are retained at their nominal values of the best fit.
The solution is integrated over $10^4$~yr for each initial condition and a stability indicator is computed to be the variation in the measured mean motion over the two consecutive 5,000~yr intervals of time. For regular motion, there is no significant variation in the mean motion along the trajectory, while it can vary significantly for \hbindex{chaotic orbits} \citep{Laskar_1990, Laskar_1993PD, Correia_etal_2005, Correia_etal_2009}.
The result is reported in color, where ``red'' represents the strongly chaotic trajectories, and ``dark blue''  the extremely stable ones.
In both plots (Figs.\,\ref{F4}a,b), there is perfect coincidence between the stable 3:2 resonant islands, and curves of minimal $\chi^2$ obtained in comparing with the observations.
Since these islands are the only stable zones in the vicinity, this picture presents a very coherent view of dynamical analysis and observational data, which reinforces the confidence that the HD\,45364 system is in a 3:2 resonant state.

\section{Formation and evolution of mean-motion resonaces}

%From a conservative point-of-view, two-planet mean-motion resonances are stable. However,
%In this section we describe the evolution of a resonant system undergoing dissipation.
During the initial stages of \hbindex{planetary formation}, planets are embedded in an \hbindex{accretion disk} and their orbits can be modified through dissipative processes.
This may lead to capture of the planets in resonance, and as long as the disk is present the orbits continue to evolve.
The two main parameters that have to be tracked during this evolution are the parameter
$\delta$ (see Eq.\,(\ref{eq:delta}))
which describes the evolution of the phase space (and of the eccentricities for resonant systems),
and the relative amplitude $A=\sin^2(\theta_\mathrm{max}/2)$ (see Fig.\,\ref{Figb}),
which describes the spiraling of the trajectory
with respect to the separatrix of the resonance.
Dissipative forces act on the semi-major axes and the eccentricities of both planets.
The evolution of the system can be described by the three following timescales
(which may depend on the semi-major axes and eccentricities of the planets):
$(\xi_1/\dot{\xi}_1)_d$, $(\xi_2/\dot{\xi}_2)_d$, and $(\alpha/\dot{\alpha})_d$, with $\alpha = a_1/a_2$.
For small eccentricities, we have $\xi_i\approx e_i$ (Eq.\,(\ref{eq:xi})), thus $(\xi_i/\dot{\xi}_i)_d \approx (e_i/\dot{e}_i)_d$.

The evolution of $\delta$ is given by \citep{Delisle_etal_2015}
\begin{eqnarray}
  \label{eq:ddeltadt}
  \dot{\delta}|_d
  &=& 2 \left(\cos^2\phi \left.\frac{\dot{\xi}_1}{\xi_1}\right|_d
    + \sin^2\phi \left.\frac{\dot{\xi}_2}{\xi_2}\right|_d + \frac{\Lambda_2 - \sin^2\phi}{4} \left.\frac{\dot{\alpha}}{\alpha}\right|_d \right) \mD   + \frac{q}{p} \frac{\Lambda_1\Lambda_2}{2} \left.\frac{\dot{\alpha}}{\alpha}\right|_d \ ,
\end{eqnarray}
while the \hbindex{libration amplitude} (for systems captured in resonance) follows
\begin{equation}
  \label{eq:dAdt}
  \left<\dot{A}\right> = \frac{1}{2 R \delta^{q/2}} \left( \left<\epsilon \dot{\epsilon}|_d\right>
    - \frac{q}{4\delta}\left<\epsilon^2\dot{\delta}|_d\right> \right),
\end{equation}
with
\begin{equation}
\epsilon = \mD - \delta \ , \quad \mathrm{and} \quad  \dot{\epsilon}|_d = - \frac{q}{p} \frac{\Lambda_1\Lambda_2}{2} \left.\frac{\dot{\alpha}}{\alpha}\right|_d \ .
\end{equation}

Because of \hbindex{disk-planet interactions}, a torque is exerted on the planets that induces a modification in their orbital elements and subsequent \hbindex{migration} in the disk
\citep[e.g.,][]{Goldreich_Tremaine_1979c, Goldreich_Tremaine_1980}.
In particular, the angular momentum of each planet evolves on an exponential migration timescale
$T_{m,i}$, while eccentricities evolve on an exponential damping timescale $T_{e,i}$
\citep[e.g.,][]{Papaloizou_Larwood_2000, Terquem_Papaloizou_2007, Goldreich_Schlichting_2014}:
\begin{equation}
  \left.\frac{\dot{\scaled{G}}_i}{\scaled{G}_i}\right|_d = - \frac{1}{T_{m,i}}
  \ , \quad \mathrm{and} \quad
  \left.\frac{\dot{e}_i}{e_i}\right|_d = - \frac{1}{T_{e,i}} \ ,
\end{equation}
which translates to
\begin{equation}
  \label{eq:diskTe}
  \left.\frac{\dot{\xi}_i}{\xi_i}\right|_d \approx \left.\frac{\dot{e}_i}{e_i}\right|_d = -\frac{1}{T_{e,i}} \ ,
\end{equation}
and
\begin{equation}
  \label{eq:diskTalpha}
  \left.\frac{\dot{\alpha}}{\alpha}\right|_d
  = \frac{2}{T_m}  + \frac{4}{\Lambda_1\Lambda_2} \left(\frac{\Lambda_1\sin^2\phi}{T_{e,2}} - \frac{\Lambda_2\cos^2\phi}{T_{e,1}} \right) \mD \ ,
\end{equation}
where $T_m^{-1} = T_{m,2}^{-1} - T_{m,1}^{-1}$.
With these simple decay laws, we can deduce the evolution of the parameters of interest
for resonant systems ($\dot{\delta}$ and $\dot{A}$).
Assuming that the damping timescale is much shorter than the migration timescale ($T_{e,i}\ll T_{m,i}$) \citep[e.g.][]{Goldreich_Tremaine_1980}, and neglecting second order terms in $\mD$ ($\mD^2 \propto e^4$), from expression (\ref{eq:ddeltadt}) we obtain \citep{Delisle_etal_2015}
\begin{equation}
  \label{eq:diskdeltamean}
  \dot{\delta}|_d = \frac{1}{T_M} - \frac{\mD}{T_E}
    \ , \quad \mathrm{and} \quad
  \left<\dot{\delta}|_d\right> = \frac{1}{T_M} - \frac{\delta}{T_E} \ ,
\end{equation}
with
\begin{equation}
  \label{eq:TEapprox}
\frac{1}{T_M} = \frac{q}{p}\frac{\Lambda_1\Lambda_2}{T_m}
  \ , \quad \mathrm{and} \quad
    \frac{1}{T_E} =  2 (\Lambda_1+\Lambda_2) \left( \frac{p+q}{p} \frac{\cos^2\phi}{T_{e,1}} + \frac{\sin^2\phi}{T_{e,2}} \right).
\end{equation}
These timescales can also be expressed using more usual notations
\begin{equation}
  \label{eq:TMTE}
  \frac{1}{T_M} \approx \frac{q}{p} \, \frac{m \alpha_0^{3/2} T_{m}^{-1}}{2 m + \alpha_0 + m^2 \alpha_0^5}
  \ , \quad \mathrm{and} \quad
  \frac{1}{T_E} \approx 2 \, \frac{1+ m \alpha_0^{1/2}}{1+ m\alpha_0^{-1}} \,
  \frac{T_{e,2}^{-1} + m e_0^2\alpha_0^{-1} T_{e,1}^{-1} }{1+ m e_0^2 \alpha_0^{1/2} } \ ,
  \end{equation}
where $\alpha_0 \approx (p/p+q)^{2/3}$ and $e_0 = (e_{1} / e_{2})_{ell} $ are respectively the semi-major axis ratio and the eccentricity ratio at the center of the resonance (Fig.\,\ref{Figb}), and $m= m_1/m_2$.

Depending on the values of $T_M$ and $T_E$, different evolution scenarios for $\delta$ are possible.
All these equations remain valid for $T_M$ and $T_E$ negative.
In most cases the disk induces a damping of eccentricities ($T_{e,i} > 0$, thus $T_E>0$),
but some studies \citep[e.g.,][]{Goldreich_Sari_2003} suggest
that an excitation of the eccentricities by the disk is possible ($T_{e,i} < 0$, thus $T_E<0$).
The timescale $T_M$ is positive if the period ratio between the planets ($P_2/P_1$) decreases (convergent migration).
But if the planets undergo \hbindex{divergent migration} ($P_2/P_1$ increases), $T_M$ is negative.
This does not depend on the absolute direction (inward or outward)
of the migration of the planets in the disk, but only on the evolution of their period ratio.

\begin{figure}
    \centering \includegraphics[width=\linewidth]{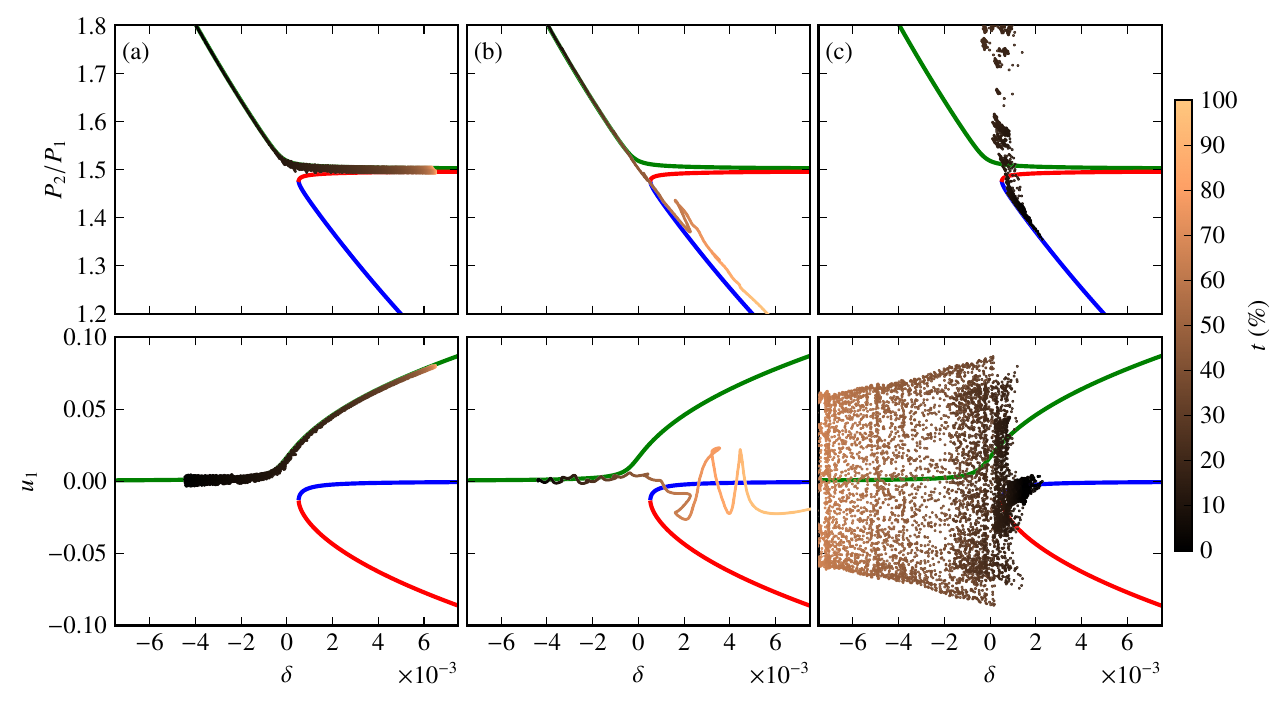}
    \caption{Tracks of three simulations in the ($\delta$,$u_1$) plane (\textit{top}),
    and ($\delta$,$P_2/P_1$) plane (\textit{bottom}),
    superimposed with the positions of the fixed points
    of the 3:2 resonance in the HD\,45364 system (see Fig.\,\ref{Figg}).
    The initial semi-major axis of the inner planet is set to 10~AU,
    both orbits are circular and the simulation start at conjunction.
    The ratio $K=T_{m,i}/T_{e,i}$ is set to 15.
    In {(a)},
    we set $a_2=15$~AU such that the period ratio ($\approx 1.8$) is higher than 3/2,
    the inner planet does not migrate, and the outer planet migrates inward
    (convergent migration)
    with a timescale $T_{m,2}=5\times 10^5$~yr.
    The system is captured in resonance following the green fixed point (see Fig.\,\ref{Figb}) and the eccentricities are excited ($u_1$ increases).
    In {(b)}, the initial conditions are the same as in {(a)}, except $T_{m,2}=5\times 10^3$~yr.
    The system crosses the resonance without beeing captured,
    and the track jumps from the green to the blue fixed point
    (with a non-negligible amplitude of oscillation).
    In {(c)}, we set $a_2=12.25$ ($P_2/P_1\approx 1.36 < 1.5$),
    the outer planet does not migrate, and the inner planet migrates inward
    (divergent migration)
    with a timescale $T_{m,1}=5\times 10^5$~yr.
    The system crosses the resonance without being captured,
    and the track jumps from the blue to the green fixed point
    (with a non-negligible amplitude of oscillation).
    }
    \label{Figsuperpos}
  \end{figure}

In the case of divergent migration, the planets cannot be trapped in resonance \citep[e.g.,][]{Henrard_Lemaitre_1983}.
The system always ends up with a period ratio higher than the resonant value, and this does not depend
on the damping or excitation of eccentricities.
The case of \hbindex{convergent migration} is more interesting.
If the initial period ratio is higher than the resonant value, the planets can be locked in resonance.
It has been shown \citep[see][]{Henrard_Lemaitre_1983, Batygin_2015}
that for first order resonances,
if the migration is slow enough (adiabatic condition),
and the initial eccentricities are small,
the capture in resonance is certain (probability of 1).
In Fig.\,\ref{Figsuperpos}, we illustrate three different scenarii of resonance crossing
for \hbindex{HD\,45364}.
In the first case (Fig.\,\ref{Figsuperpos}\,{(a)}),
the migration is convergent and the planets get captured in resonance.
The period ratio is locked at the resonance value
(Fig.\,\ref{Figsuperpos}\,{(a)} {\it top}),
while the eccentricities (or equivalently $|u_1|$) increase
(Fig.\,\ref{Figsuperpos}\,{(a)} {\it bottom}).
%until the system reaches the equilibrium value $\delta_{eq}\approx 6\times 10^{-3}$.
In the second case (Fig.\,\ref{Figsuperpos}\,{(b)}),
the migration is convergent but much faster,
which causes the resonance crossing to be non-adiabatic
\citep{Henrard_Lemaitre_1983,Batygin_2015}.
The system crosses the resonance without being locked
(Fig.\,\ref{Figsuperpos}\,{(b)} {\it top}),
and the excentricities are slighlty excited by this crossing
(Fig.\,\ref{Figsuperpos}\,{(b)} {\it bottom}).
Finally, in the third case (Fig.\,\ref{Figsuperpos}\,{(c)}),
the migration is divergent,
and the system crosses the resonance without being locked
(Fig.\,\ref{Figsuperpos}\,{(c)} {\it top}).
The excentricities are significantly excited by the resonance crossing
(Fig.\,\ref{Figsuperpos}\,{(c)} {\it bottom}).

The capture in resonance induces an excitation of the eccentricities of the planets
($\dot\delta|_M = 1/T_M> 0$).
If $T_E<0$ (excitation of the eccentricities by the disk) or $T_E \gg T_M$ (inefficient damping),
$\delta$ (as well as the eccentricities) does not stop increasing.
When the eccentricities reach values that are too high,
the system becomes unstable, and the resonant configuration is broken.
However, the most common scenario is the case of efficient damping of the eccentricities ($0 < T_E \lesssim T_M$).
In this case, $\delta$ reaches an equilibrium value, for which the eccentricities stabilize ($\langle\dot\delta|_d\rangle = 0$, see Eq.~(\ref{eq:diskdeltamean})):
\begin{equation}
  \delta_{eq} = \frac{T_E}{T_M} \ .  \label{eq:diskdeltaeq}
\end{equation}
%However, as shown by \citet{Goldreich_Schlichting_2014}, for the restricted three body problem, this equilibrium can be unstable.

In the case of \hbindex{convergent migration},
even if the system is captured in resonance and the parameter $\delta$
reaches the equilibrium $\delta_{eq}$,
the system can still evolve.
Indeed, the libration amplitude can increase
until the system crosses the separatrix and escapes from resonance.
According to Eq.~(\ref{eq:dAdt}), the evolution of the amplitude depends on
$\langle\epsilon \dot{\epsilon}|_d\rangle$ and  $\langle\epsilon^2\dot{\delta}|_d\rangle$.
We have
\begin{eqnarray}
  \label{eq:diskA1}
  \left<\epsilon \dot{\epsilon}|_d\right> &=&
  2 \frac{q}{p} \left(\frac{\Lambda_2\cos^2\phi}{T_{e,1}} -\frac{\Lambda_1\sin^2\phi}{T_{e,2}}\right) \left<\epsilon^2\right>,\\
  \label{eq:diskA2}
  \left<\epsilon^2\dot{\delta}|_d\right> &=& \left(\frac{1}{T_M} - \frac{\delta}{T_E} \right) \left<\epsilon^2\right>
  = \left<\dot{\delta}|_d\right>\left<\epsilon^2\right>,
\end{eqnarray}
where $\langle\epsilon^2\rangle$ can be computed using elliptic integrals \citep[see][]{Delisle_etal_2014}
\begin{equation}
  \label{eq:tidePlaA3}
  \left<\epsilon^2\right> \approx 2 R \delta^{q/2} A.
\end{equation}

The first term ($\langle\epsilon \dot{\epsilon}|_d\rangle$) does not depend on the migration timescale but only on the damping timescale.
The second term ($\langle\epsilon^2\dot{\delta}|_d\rangle$) vanishes
when the system reaches the equilibrium $\delta = \delta_{eq}$, since $\langle\dot\delta|_d\rangle=0$.
This is not surprising because the first term describes the evolution of the absolute libration amplitude $\epsilon^2$, while the second one describes the evolution of the resonance width, which does not evolve if
the phase space does not evolve (constant $\delta$).
Finally, we obtain (see Eq.~(\ref{eq:dAdt}))
\begin{equation}
  \label{eq:diskAtot}
  \left.\frac{\dot{A}}{A}\right|_d =
  2 \frac{q}{p} \left(\frac{\Lambda_2\cos^2\phi}{T_{e,1}} - \frac{\Lambda_1\sin^2\phi}{T_{e,2}}\right).
\end{equation}
The libration amplitude increases if
\begin{equation}
  \frac{\Lambda_2\cos^2\phi}{T_{e,1}} > \frac{\Lambda_1\sin^2\phi}{T_{e,2}}
  \quad \Leftrightarrow \quad
  \frac{T_{e,1}}{T_{e,2}} < \frac{}{}\left(\frac{\xi_1}{\xi_2}\right)_{ell}^2 \approx \left(\frac{e_1}{e_2}\right)_{ell}^2 \ ,
  \label{eq:diskCritA}
\end{equation}
where the eccentricity ratio is evaluated at the elliptical fixed point ($ell$ subscript) at the center of the resonance.
In the outer circular restricted case
$e_2 = 0$ and $T_{e,2} = +\infty$, so the amplitude always increases %(Eq.~(\ref{eq:diskCritA}))
and the equilibrium is unstable \citep{Goldreich_Schlichting_2014}.
However, in the inner circular restricted case ($e_1 = 0$ and $T_{e,1} = +\infty$),
 the amplitude always decreases %(Eq.~(\ref{eq:diskCritA}))
 leading to a stable equilibrium \citep{Delisle_etal_2015}.
This result assumes that the eccentricity ratio
remains close to the forced value (value at the elliptical fixed point).
This is verified for a small libration amplitude, but when the
amplitude increases, the eccentricity ratio oscillates and may differ significantly from the
forced value.
This model thus only provides a first approximation of the mean value of this ratio
in the case of a high libration amplitude.

When a system is observed to host two planets locked in a mean-motion resonance,
such as HD\,45364,
it is likely that this resonant configuration was stable (or unstable but with a very long timescale) when the disk was present.
%We could imagine that the configuration was highly unstable, but the protoplanetary disk disappeared before the system had time to escape from resonance. However, this would require a fine tuning of the disk disappearing timing.
Thus, the present eccentricities should still correspond to the equilibrium ones (\ref{eq:diskdeltaeq}), and the libration amplitude was probably either decreasing or increasing on a very long timescale (Eq.~(\ref{eq:diskCritA})):
\begin{equation}
  \frac{T_{e,1}}{T_{e,2}} \gtrsim \left(\frac{e_1}{e_2}\right)_{ell}^2 \ .
  \label{eq:constBigAbis}
\end{equation}
The ratio $T_E/T_M$ of the damping and migration timescales is then constrained by the observations.
This ratio depends on the four timescales involved in the model, which themselves depend on the properties of the disk and the planets.
There is a wide diversity of disk models in the literature, but constrains on the primordial disk can be easily obtained  as long as expressions for $T_{e,1}$, $T_{e,2}$, $T_{m,1}$, and $T_{m,2}$ are available.
For instance, for \hbindex{type-I migration} we can write \citep{Delisle_etal_2015}:
\begin{equation}
  \label{eq:deftau}
  \tau \approx \frac{T_{m,1}}{T_{m,2}} \approx \frac{T_{e,1}}{T_{e,2}}
  \ , \quad \mathrm{and} \quad
  K \approx \frac{T_{m,1}}{T_{e,1}} \approx \frac{T_{m,2}}{T_{e,2}} \ ,
\end{equation}
where $\tau$ and $K$ are parameters linked to the disk local surface density and scale high, respectively.
The constraint provided by the observation of the equilibrium eccentricities reads
as (see Eq.~(\ref{eq:diskdeltaeq}))
\begin{equation}
  \label{eq:constK}
  K = \frac{C_1}{\delta} \frac{\tau-1}{\tau + C_2},
\end{equation}
where $\delta$ (Eq.\,\ref{eq:delta})),
\begin{equation}
  C_1 = \frac{1}{2} \frac{1+m e_0^2 \alpha_0^{1/2}}{1 + \frac{p}{q} ( m \alpha_0^{-1} +
     m^{-1} \alpha_0^{-1/2} ) }
  \ , \quad \mathrm{and} \quad
        C_2 = m e_0^2 \alpha_0^{-1} \ ,
\end{equation}
can all be derived from the observations.
Since $K$ is an increasing function of $\tau$, the analytical criterion for stability (Eq.\,(\ref{eq:constBigAbis})) provides a lower bound for both $\tau$ and $K$.
We apply these analytical criteria to the \hbindex{HD\,45364} planetary system, which is observed in a 3:2 resonance \citep{Correia_etal_2009}.
We have for this system $\delta = 6.0\times 10^{-3}$, $C_1 = 0.93$, and $C_2 = 2.3$ with a forced eccentricity ratio $e_0 \approx 2.5$.
The stability constraints give
$ \tau \gtrsim  6.3 $ (Eq.~(\ref{eq:constBigAbis})) and $ K \gtrsim 9.4$  (Eq.~(\ref{eq:deftau})).

\begin{figure}
    \includegraphics[width=\linewidth]{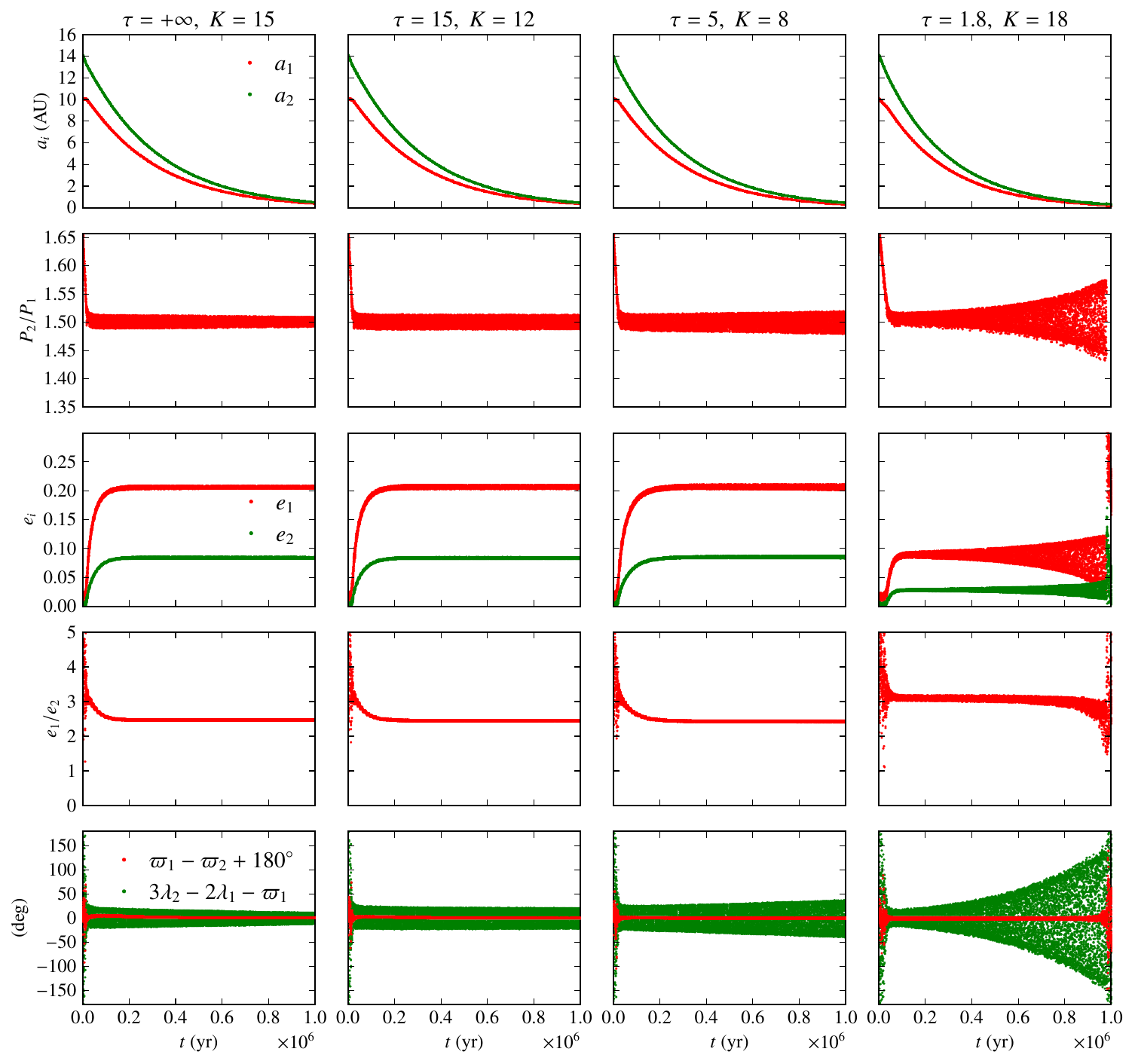}
    \caption{Semi-major axes, period ratio, eccentricities, eccentricity ratio,
      and angles evolution for the HD\,45364 system
      with different dissipation timescale ratios $\tau=T_{e,1}/T_{e,2}=T_{m,1}/T_{m,2}$.
      The ratio $K=T_{m,i}/T_{e,i}$ is set according to expression (\ref{eq:constK}) to
      reproduce the observed equilibrium eccentricities (except for the last column).
      We used $\tau = +\infty$, 15, 5, and 1.8 with K = 15, 12, 8, and 18 respectively,
      for the four shown simulations (four columns).
      The libration amplitude decreases for the first two simulations
      ($\tau=+\infty$, 15) and increases for the last two ($\tau=5$, 1.8).
      The value given by our analytical criterion for the transition
      between decreasing and increasing amplitude is $\tau\sim 6.3$.
      In the last case (\textit{right}),
      the system escapes from the resonance just before the end of the 1~Myr simulation.}
    \label{fig:IV}
  \end{figure}

We also performed numerical simulations
with different values of $\tau$
and $K$ (given by Eq.~(\ref{eq:constK})).
We fixed the migration timescale for the outer planet ($T_{m,2}=5\times 10^5$~yr),
and varied the three other timescales: $T_{m,1} = \tau T_{m,2}$,
$T_{e,2} = T_{m,2}/K$,
$T_{e,1} = \tau T_{m,2}/K$.
The semi-major axes are initially 10 and 14~AU (period ratio of about 1.65),
the eccentricities are 0.002 with anti-aligned periastrons and
coplanar orbits.
The planets are initially at periastrons (zero anomalies).
The evolution of the semi-major axes, the period ratio, the eccentricities,
the eccentricity ratio, and the angles are shown in Fig.~\ref{fig:IV}.
According to the simulations, the libration amplitude increases
for $\tau \lesssim 10$ (transition between 5-15, see Fig.~\ref{fig:IV}),
which is comparable to the analytical result ($\tau\lesssim 6.3$).
% It may seem surprising that the libration amplitude does not increase much more rapidly
% for $\tau=2$ than for $\tau=5$ (see Fig.~\ref{fig:IV}).
% Indeed, the smaller $\tau$ is,
% the more unstable the resonant configuration is (Eq.~(\ref{eq:diskAtot})).
% However, the evolution of the libration amplitude not only depends on the
% ratio $\tau = T_{e,1}/T_{e,2}$, but also on the absolute values of
% these damping timescales (see Eq.~(\ref{eq:diskAtot})).
% The damping timescales are thus much longer for the simulation with
% $\tau=2$ ($K=3.5$) than for $\tau=5$ ($K=8$), in order to reproduce the
% same equilibrium eccentricities.
% This tends to slow down the increase in the libration amplitude
% and compensates for the acceleration provided by decreasing $\tau$.

\section{Conclusion}
\label{sec:conclusion}

We have revisited here the dynamics of mean-motion resonances, their detection, stability, formation and evolution.
Due to the strong mutual gravitational interactions between planets involved in resonances, the dynamics is more complex than for non-resonant systems.
However, when we take into account some simplifying assumptions, we can reduce the degrees-of-freedom, and the resonant dynamics can be described by the level curves of an integrable problem (Figs.~\ref{Figb} and \ref{Figg}).
Since the orbital elements that we obtain from exoplanet observations are still subject to some uncertainties, it is not straightforward to determine whether
a system is in resonance or not.
In particular, the simple computation of the period ratio of two planets is usually not sufficient to decide if they are in resonance. 
A system can only be said to be in a mean-motion resonance with confidence when it is observed in libration in between the separatrices of the resonance (Fig.\,\ref{Figb}),
and that no stable orbit outside of the resonant island is compatible with the observations (Fig.\,\ref{F4}).
The evolution of a resonant pair of planets undergoing disk-planet interactions mainly depends on two parameters:
$T_E/T_M$ (damping vs migration timescale) and $T_{e,1}/T_{e,2}$ (damping in inner planet vs outer planet).
The ratio $T_E/T_M$ governs the equilibrium eccentricities of the planets
(Eqs.~(\ref{eq:diskdeltamean}) and (\ref{eq:diskdeltaeq})).
The ratio $T_{e,1}/T_{e,2}$ governs the stability of this equilibrium
(Eqs.~(\ref{eq:diskAtot}) and (\ref{eq:diskCritA})).
The present resonant configuration thus allow us to put constraints on these timescales, which in turn are related to the surface density and scale high of the initial disk.
In this paper we have selected one real example, the HD\,45364 planetary system, to illustrate all previous results.
This system has two giant planets involved in a 3:2 mean-motion resonance, whose orbits are very close and stability can only exist due to the presence of the protective resonant mechanism.
Other systems exist (see Table~\ref{tab1}), with different masses and orbital period ratios, but the global picture drawn for HD\,45364 should not vary much.

\begin{acknowledgement}
The authors acknowledge financial support from the Observatoire de Paris Scientific Council, CIDMA strategic project UID/MAT/04106/2013, and Programme National de Planetology,  SNSF.
This work has, in part, been carried out within the framework of
the National Centre for Competence in Research PlanetS
supported by the Swiss National Science Foundation.
\end{acknowledgement}

%  IF you do NOT use bibtex, put comments before the following 2 lines
\bibliographystyle{spbasicHBexo}  %for bibtex
\bibliography{correia} %for bibtex-example

\end{document}